\documentclass[namedreferences]{kluwer}

\usepackage{float}
\usepackage{amsmath,amsthm}
\usepackage[dvips]{graphicx}
\usepackage{latexsym}
\begin{document}
\begin{article}
\newtheorem{theorem}{Theorem}
\newtheorem{lemma}{Lemma}
\newtheorem{corollary}{Corollary}
\newcommand {\rel}{\,{}^{{}_\simeq}_{{}^\sim}}

\begin{opening}
\title{Set Theoretical Forcing in Quantum Mechanics and AdS/CFT correspondence.}
\author{{Jerzy} \surname{Kr\'ol}\footnote{Institute of Physics, University of Silesia, Uniwersytecka 4, 40-007 Katowice, Poland }\footnote{\email{iriking@poczta.fm}}}
\institute{} 

\begin{abstract}
We show unexpected connection of Set Theoretical Forcing 
with Quantum Mechanical 
lattice of projections over some separable Hilbert space. The basic ingredient of the construction is the rule of indistinguishability of Standard 
and some Nonstandard models of Peano Arithmetic. The ingeneric reals introduced by M. Ozawa will correspond to simultaneous measurement of incompatible observables. 
We also discuss some results concerning model theoretical analysis of Small Exotic Smooth Structures on topological 4-space $\mathbb{R}^4$. Forcing appears rather naturally in this context and the rule of indistinguishability is crucial again. As an unexpected application we are able to approach Maldacena Conjecture on $AdS/CFT$ correspondence in the case of $AdS_5\times S^5$ and Super YM Conformal Field Theory in 4 dimensions. We conjecture that there is possibility of breaking Supersymetry via sources of gravity generated in 4 dimensions by exotic smooth structures on $\mathbb{R}^4$ emerging in this context.   
\end{abstract}

\end{opening}

\section{Remarks on Model Theory and QM.}
Since the very beginning of Model Theory (MT), it was realized what is the position of formal language in the correct description of any Mathematical structure. There is a big difference when we talk about same things using different formal languages. Many "paradoxical" situations emerged in the development of the so called first order logic (let me mention only L\" owenheim-Skolem Theorem or G\"odel Theorems), which nonetheless serves as the logic depending on very minimal theoretical support from Set Theory and the like. It has resulted in emergence of various spectra of non isomorphic models, which first order theories must have (\opencite{KeislerChang}). One can say, that these are merely formal peculiarities, which are not acting upon our every day practice (espacially as a physicists). Surprisingly enough, the theories such as: Peano Arithmetic (PA), or axiomatized theory of natural numbers and Zermello-Fraenkel Set Theory (with the Axiom of Choice [AC]) ZF(C), or axiomatized theory of sets---both are first order theories. This means that all paradoxical situations concern also them. Because of the fact that very effective formalism of Quantum Mechanics (QM) was developed very early, people were not considering tools of MT in the context of QM (also, they simply were not able to do so, because MT was being developed in the time Quantum Theory was as a young branch of Mathematics). In the paradigm of QM people found it possible to talk about Many Worlds (as in the Everett interpretation of QM), but no one has tried to incorporate Model Theoretical tools in the shape of pluralities of models of basic enough theory as ZFC is. In what follows we are giving some insights coming from MT, which might be helpful in analysing certain interpretational and formal aspects of Quantum Theory.

Firstly, in the face of the L\"owenheim-Skolem Theorem (\opencite{KeislerChang}) it follows that ZFC (if it is consistent, \inlinecite{Jech}) has a countable model. In particular, the set of real numbers in this model is countable from the "outside" (\opencite{Jech}). Might it be, that discreteness of spectra of some physical Quantum Mechanical observables is connected to the discreteness of real numbers in some models of ZFC?

Secondly, might it be that pluralities and non isomorphism of models of ZF(C) (or some other formal theories) can be connected to pluralities of Quantum Mechanical worlds in Everettian style?

Thirdly, as follows from Kochen - Specker theorem (\opencite{KSpecker}) and EPR kind of experiments, some physical observables cannot have definite values before measurements. Might this phenomenon be connected to nonexistence of some real numbers in some models of ZF(C)? Might it be that the real numbers in question, which are not present in one model, will appear in the other as a result of Set Theoretical forcing? That would mean that the reals obtained as a result of measuring of some observables would emerge as a result of being forced to belong to some enlarged universum of sets. In this situation it would mean that there are inherent reasons for considering a non- constant universum of sets in the context of QM.

Whereas we avoid deciding about the import of the first two situations, we are to explore the third one. In general, detailed proofs are not included here. Informal, intuitive explanations are often used in place of them. The proofs will be given elsewhere (\opencite{Krol2}). 

We assume some elementary knowledge of Set Theoretical forcing; the general reference is \inlinecite{Jech} which also serves as a source for all set theoretical questions considered here. Some knowledge of Boolean Valued models of ZFC is also desirable; the general reference is \inlinecite{Bell1985+}. In the context of QM, Boolean Valued models are discussed in Takeuti \shortcite{Takeuti1978, Takeuti1979, Takeuti1983}.
General references concerning Topos Theory are \inlinecite{Johnstone} and \inlinecite{Moerdijk}.

There were some attempts recently, to interpret QM in terms of Topos Theory: \inlinecite{Isham}, \inlinecite{Fearins} (by the use of so called Smooth Toposes from Synthetic Differential Geometry (SDG) (\opencite{ReyesMoerdijk})) or \inlinecite{Raptis} in the context of Quantum Gravity. Also, there are some attempts to approach QM through Model Theoretic tools (\inlinecite{Benioff}, \inlinecite{Krol}, Takeuti \shortcite{Takeuti1978,Takeuti1979}). Much effort has been spent on working out connections of QM (and other branches of physics) with Nonstandard Analysis (\opencite{NA}) invented by \inlinecite{Robinson} as a purely Model Theoretical task (Robinson himself was one of the creators and main contributors to Model Theory). Close connections of QM and formal logic were noticed from the very beginning of the subject (\opencite{Neumann}). The so called Quantum Logic (QL) emerged as a special one and different from Classical Logic and Intuitionistic Logic (\opencite{QL}).

Synthetic Differential Geometry (SDG) was worked out by \inlinecite{Kock} and then developed by \inlinecite{ReyesMoerdijk}. The aim of this approach was to build common framework for Toposes and Nonstandard Analysis in the Robinsonnian style, as well as for differential geometry, which deals with infinitesimals called noninvertible or indempotent. Needless to say, such unification is of great interest to physicists, too. Because of that, there were some trials to use these techniques in the context of Quantum and Classical Gravity (\opencite{Guth}) or in QM (\opencite{Fearins}). Very nice features of Toposes, which are inducing Intuitionistic Logic, and are natural models for arbitrarily high order logic, cause people naturally start to think about using the strangeness of the world generated by Toposes (the Midle Third Law can be violated in Toposes, Standard Natural Numbers may be not a decidable object, there may not exist an object of naturals at all, Set Theory is generated in the inside of the Topos; Toposes might serve as Universes of Discourse for Mathematics, and so on (\opencite{Moerdijk})) to the strangeness (at the interpretational level) of the Quantum World. However, one should be aware that:

{\bf Warning:} {\it No Topos is able to interprete QM in the sense that the Logic generated by the Topos is the one generated by Quantum Mechanical phenomena.}

Proof: Logical structure of any Topos is based on Heyting Algebras (HA). Any HA is distributive one. Quantum logic is based on the Lattice of projections (\opencite{Bell1985}) in some separable Hilbert space. Such a Lattice is in general non distributive (\opencite{RS}, \opencite{Takeuti1978}) (the lattice of projections is distributive only in the case of one dimensional Hilbert space).  $\Box$

\section{Forcing, Set Theory and QM.}
Forcing was discovered originally as a tool for proving some
independence results in Set Theory from the axioms of ZF(C)
(\opencite{Cohen1963}) and then, developed as very basic phenomenon in
many different mathematical situations. Solovay and 
\inlinecite{Scott} and, independently, \inlinecite{Vopenka} have given
formulation of forcing in the language of so called Boolean valued
universes of ZF(C). Forcing as it stands has been shown to have
specific significance in very general situations in the Model Theory
(\opencite{RobinsonBarwise}). Semantics connected to
Toposes (at least to Grothendieck Toposes) in a natural way is also modeled on
forcing semantics (sheaf semantics) (\opencite{Johnstone}). The variety
of  ways the situation of forcing appears in many different levels
of mathematical reasoning (proof theory, logic, topos semantic,
model theory) causes it to seem to be really fundamental phenomenon.

Forcing, despite its formal shape, has very specific action over universes of Set Theory (ZF(C)): it defines the relations between names of things, which are not present in the Universe (it means they are not sets in this sense) such that the relations hold between sets (whose names were considered) in the enlarged Universe. The procedure of enlargement with the preservation of theorems of ZF(C), is just the forcing procedure. From the other perspective forcing is the recipe for proving theorems valid in enlarged Universe (which is a Model of ZF(C)) and not necessarily valid in the ground Universe (which is also a Model of ZF(C)). In that way Cohen was able to proove independence of Continuum Hypothesis (CH) and the Axiom of Choice (AC) from the axioms of  ZF.

\inlinecite{Bell1983} considered the situation of forcing in some approach to Quantum Logic. He was able to point out the place in Quantum Logic, where forcing, understood logically, fails to work. 
There are some indications coming from his analysis that:

(i)  Set theoretical context is probably changing.

(ii)  There are some "preformal" situations where nonstandard model of reals (in the Robinsonnian sense) i.e. ${^\ast}{\bf R}$ and standard one -- $\mathbb{R}$, are "identified" (or, in fact, ${\mathbb{N}}$ and ${^\ast}{\bf N}$ as models of Peano Arithmetic (PA) are identified).\\
In what follows we are to investigate both points. Strikingly enough, this will lead us to rebirth of forcing in the context of QM and, in fact, forcing appears to be very intrinsic shape of Quantum Mechanical paradigm; instead of being only formal construction in logic, or QL, it is to be considered as a tool allowing the change of Metatheoretical Universe and/or the way reals appear as possible results of experiments. In fact, both points (i) and (ii) intertwine in the specific way, which is  well modeled locally by the model--theoretical constructions of Bounded Boolean Ultrasheaf of Superstructure modulo some Ultrafilter $(bBUSuS/U)$ in Boolean Algebra (BA) (invented by \inlinecite{Ozawa1994})(see Theorem3). \\
In this paper we show surprising correlation of formalism of QM with some structures which are modelled on forcing constructions in various of its facets. This is still more surprising, because it was shown by \inlinecite{Bell1985} that the forcing formally does not work any more in QL (what we have mentioned already). The missing link for appearing forcing in QM formalism could be non classical $Main$ $Hypo$.

\section{The Rule of Indistinguishability Standard and Some
Nonstandard Models of Peano Arithmetic.}
 
The rule is exactly as stated in the title, but it requires some
explanation. Firstly, the Rule cannot be formalized; otherwise it
would exist formal context where one could express: Standard
and Nonstandard models of PA are different and (according to the Rule) one cannot
express their difference---this is impossible. For, there does exist
suitably higher order formal language (in fact second order is enough), which expresses the difference between Standard N and Nonstandard N; there would also
exist an order, and the language of that order, which would express identity of both models. It is enough to take the language of order being maximal of both orders (with appropriate symbols), to obtain contradiction.

Secondly, the Rule is the heuristic statement only; one cannot prove it (this would require some formal context).\\ 
In what follows, we give some possible formulations of the Rule.

{\it In some formal constructions (not necessarily connected with
Nonstandard Naturals) the appearance of Nonstandard Naturals is
unavoidable and out of control by formal means of the construction.}

{\it Along with the shift of the language $L$ from the higher order into first order, we cannot distinguish Standard N from Nonstandard N. There are such shifts which are not able to be described simultaneously in any formal language, which would be an extension of the language $L$.}

{\it For any given formal language (of any order) there exists a pair $(\mathbb{N}, {\bf ^\ast N})$ such that there doesn't exist any formula $\phi(\vec{x})$ (with free variables from $\vec{x}$) in this language, which would be expressing any difference between  $\mathbb{N}$ and ${\bf ^\ast N}$.}

An analogy with Manifolds can be usefull here: we have patches which are Standard N or Nonstandard N, but the transition functions between them cannot be formalized simultaneously with pathes (the language--logic is also to be shifted). Such a thing we call {\it Levelfold}.

As we will see, such a strange Rule, called from now on $Main$ $Hypo$, has much to do with formalism of QM.

{\it There does not exist any formal language which would be able both to describe language of PA (counting formulas of it) and to describe {\bf all} models of PA (There are some models of PA which cannot be discribed simultaneously with the formal language of PA).}

In general one can say that non simultaneous (in principle) description of formal language and the models of the theory in this language (which in fact is a core of $Main$ $Hypo$) is analoguos to non simultaneuos (also in principle) measurability of some physical observables. We will see what exactly is the formal content of this coincidence.

\section{Forcing Constructions.}

We are ready to formulate some strange consequences of the Main Hypo.
There is canonical bijection between
morphisms of ultrafilters on $\omega $ and elementary embeddings
of ultrapowers of $\mathbb{N}$. Morphisms of ultrafilters induce the
order, which is just Rudin-Keisler (RK) order. E-morphism (End
extension morphism) is induced by being an initial segment of one
ultrapower of $\mathbb{N}$ in another. E-morpphism also induce E-order
on ultrafilters on $\omega $. This is stronger preorder then RK.
(Another possible order is the order connected to conservative
extensions of ultraproducts; it is stronger then E-order. The
strongest one is Rudin-Frolik order depicted as $\sqsubseteq $
(\opencite{Blass1977}). \inlinecite{Murakami1999} was able to define RK order and RF order on
ultrafilters in arbitrarily complete BA).
What follows from $Rudin - Keisler$ order of ultrafilters over
$P(\omega)$, taking not $Rudin -
Keisler$ related ultrafilters we are led to {\bf not end
extensions} of Nonstandard Naturals. Such an extension is to
modify some initial segment of naturals (Nonstandard):
$[0,1,2,...\bar{n}] \subset {\bf^\ast N}$ in the sense that there
is no 1 to 1 mapping $\phi$ :
\begin{displaymath}
\phi : [0,1,...\bar{n}]\not \xrightarrow[into]{1:1} [0',1',2',...,\bar{n}']
\end{displaymath}
such that $\phi (n)= n'$. They both are end extensions of Standard Naturals: they have order type $\omega + (\omega+\omega^\ast)*\eta$ , where $\eta$ is dense order without endpoints and $\omega^\ast$ is reverse order of $\omega$ (\opencite{Kamo}).
\begin{lemma}
Under hypothesis of non distinguishing of Standard N and some Nonstandard N (Main Hypo) it would be possible to have not end extensions of Standard Natural Numbers
\end{lemma}
{\it Remark:} Because all extensions of PA are end extensions of initial segments of Standard Naturals (\opencite{Blass1977}), (although they could be not conservative extensions) it is clear that the above construction cannot be fully classical (see discussion in Section 3).

\begin{lemma}
Under Main Hypo, one can have not end extension of Standard Naturals where some $k\in \mathbb{N}$ is modified into some infinite non standard natural from ${\bf ^\ast N} - \mathbb{N}$.
\end{lemma}

Next, consider lattice of projections $\cal L$ (self adjoint and indempotent operators) over some separable Hilbert space $\cal H$.
Let $(B_\alpha)$ be some maximal BA of projections from $\cal L$. 

\begin{lemma}
$(B_\alpha)$ is complete atomic Boolean Algebra.
\end{lemma}

\begin{corollary}
Maximal algebra of projections in the lattice of projections over separable Hilbert space cannot serve as a Forcing algebra for any nontrivial forcing over any Ground Model of ZF(C).
\end{corollary}

This is well known fact that forcing is not trivial (properly extends the ground model) iff the BA which corresponds to it is atomless BA (\opencite{Jech}).   

Let us notice that atomic, complete BAs fulfill infinite distributive law (DL):
\begin{displaymath}
\prod_{i \in I} \sum _{j \in J} p(i,j)= \sum_{f \in {J^I}}\prod_{i \in I} p(i,f(i))
\end{displaymath}
where $I$, $J$ are arbitrarily, possibly infinite, sets of indices and $\{p(i,j)\}_{i \in I,j \in J}$ is any, double indexed family of elements of BA.

Any non distributive elements $A$, $B$, $C$ from the lattice $\cal L$ (in the sense, that their spectral families are in $\cal L$):

(**)   $A \wedge (B \vee C) \neq A \wedge B \vee A \wedge C$ \\
cannot be simultaneously in any $(B_\alpha )$. For, if not, lets consider spectral decompositions of $A$, $B$, $C$, from (**) it follows that some triples of projections from the decompositions would also fulfill (**). That means they cannot be simultaneously in any $(B_\alpha)$.
\begin{theorem}
(1)Under Main Hypo, there is natural correspondence between lattice of projections $\cal L$ and some family of not $(\omega ,\omega)$ distributive complete Boolean Algebras.

(2)Any such Boolean Algebra can be Cohen forcing BA which adds some Cohen generic reals into some Ground model ${\bf V}$ of ZFC.

(3)Under $Main$ $Hypo$ this real interpreted in Boolean Valued model of ZFC -- ${\bf V}^{\bar{B}}$, corresponds to the self adjoint operator which is not in $(B_\alpha)$.
\end{theorem}

\begin{theorem}
Cohen generic reals from Theorem1 are able to be interpreted as a probability distributions over spectrum of Boolean Algebra of Projections $(B_\alpha )$ in the separable Hilbert space  $\cal H$. The reals might be obtained in the experiments of measuring an observable, corresponding to the self adjoint operator from Theorem1.3.
\end{theorem}
{\it Remark:} The peculiarity of Theorem2 is in that even in the single measurement having got single "real" as a result, we already have corresponding probability distribution coming from the results of repeating measurements (to be performed). The dynamics of shifts of ZFC Models is giving this situation, which in turn enables one to "realize" $Main$ $Hypo$.

Now we are ready to analyze what meaning can be given to reals obtained in measuring $A$ or $B$ (self adjoint operators). The related question is: what is the spectrum of $C_\omega$, having fixed the spectrum of $(B_\alpha)$. Both should be composed of real numbers, but once we are fixing the meaning of reals corresponding to $(B_\alpha)$, then what are the reals corresponding to $C_\omega$(interprated, by the virtue of $Main$ $Hypo$, in terms of projections from $\cal L$)? Finally, what does it mean that one cannot measure non commuting $A$ and $B$ simultaneously?.

Let us notice that reals $R^{(B_\alpha )}$ in ${\bf V}^{(B_\alpha)}$ and $R^{((B_{\alpha})+(A_s))}=R^{C_\omega}$ in ${\bf V}^{C_\omega}$ differ due Cohen generic reals on the level of ground model ${\bf V}$. This means that we have to change the ground model along with the shift $(B_\alpha) \rightarrow ((B_{\alpha})+(A_s))$, where both algebras are treated as algebras of projections from $\cal L$. That means in turn, that to maintain indistinguishability of algebras, we have changed the meaning of the Lattice of Projections. Thus we have reached the point where everything has changed the meaning (even standard Natural Numbers from $Main$ $Hypo$).

To try to see the difference between varying Naturals or Reals we trying to fix $\mathbb{N}$ and $\mathbb{R}$ and are going to employ the technique of Bounded Boolean Ultrasheaf of Superstructure Modulo some Ultrafilter in BA ($bBUSuS/U$) (\opencite{Ozawa1994}). 
$bBUSuS/U$ is in fact based on generalization of Ultraproduct modulo Ultrafilter in $P(\omega)$ (\opencite{KeislerChang}) into Boolean Ultraproduct modulo Ultrafilter in BA in question (\opencite{Mansfield}). The next step is to construct Superstructure to generate nonstandard elements inside Boolean Valued Universum of ZFC, and to make forcing, and to generate infinitesimals simultanouseously by the same Ultrafilter in BA (\opencite{Ozawa1994}).

On the base of fixed $\mathbb{R}$ we will be able to see what reals should look like in the presence of non compatible observables. Let us consider some superstructure allowing one to deal with $\mathbb{N}$, ${\bf ^\ast N}$, $\mathbb{R}$ and ${\bf ^\ast R}$ according to some Ultrafilter on $P(\omega)$ i.e. $V({\bf R}\cup P(\omega))$. On the side of "true" (connected to Standard N being well distinguished to any Nonstandard N) Lattice of Projections, maximal BA of projections in $\cal L$ is isomorphic to $P(\omega)$, (Lemma3). Following the arguments in \inlinecite{Ozawa1994} we can define superstructure in Boolean valued model ${\bf V}^{P(\omega)}\simeq{\bf V}^{(B_\alpha)}$ (being superstructure is expressible by $\triangle _0$- formulas). This means that our superstructure generates ${\bf ^\ast N}$ and ${\bf ^\ast R}$ which are equally generated inside ${\bf V}^{(B_\alpha)}$.

By the shift $(B_\alpha)\rightarrow((B_{\alpha})+(A_s))\simeq C_\omega $ (coming from $Main$ $Hypo$), we are faced with Boolean valued model ${\bf V}^{C_\omega}$ and ingeneric reals are appearing with respect to the superstructure $V(R\cup C_\omega)$ (\opencite{Ozawa1994}). Now, there is nonequivalence between ${\bf ^\ast N}$, ${\bf ^\ast R}$ generated in the outside and inside of the ${\bf V}^{C_\omega}$. More exactly, an effect of forcing corresponds inside ${\bf V}^{C_\omega}$ to ingeneric reals with respect to fixed ${\bf R}$. That means, $Main$ $Hypo$ through the shift ${\bf V}^{P(\omega)}\simeq{\bf V}^{(B_\alpha)}\rightarrow {\bf V}^{C_\omega}$ is giving ingeneric reals as those which correspond to enlargement $(B_\alpha)\rightarrow ((B_{\alpha})+(A_s))$, so ingeneric reals correspond to $A$($A$ determines generic Cohen real, which correspond to some ultrafilter $U$ in $C_\omega$. $C_\omega$ and $U$ in turn, give ingeneric reals (\opencite{Ozawa1994})). Let us remind that by a spectrum of Boolean Algebra of projections we mean the spectrum of $C^{\star}$ - algebra, which is uniquelly determined by the algebra of projections (\opencite{Jech1984}). Now we see that fixing the meaning of true {\bf R} as connected to the spectrum of $(B_\alpha)$ we can regard reals connected to the shifted algebra $((B_{\alpha})+(A_s))$ as ingeneric enlargement coming from $bBUSuS/U$ with respect to $C_\omega$ and ultrafilters in it. We have obtained the characterization of the spectrum of enlarged algebra coming from non compatible observables (through $Main$ $Hypo$).

\begin{theorem}
In the measurement of non compatible observables as a result of the possible simultaneous measurement we would have real number and some ingeneric real corresponding to Bounded Boolean Ultrasheaf of SuperStructure modulo some Ultrafilter in BA --- $bBUSuS/U$. That is why having both reals is not possible.
\end{theorem}

\begin{theorem}
According to the shift $(B_\alpha )\rightarrow ((B_\alpha )+A)$ we have the shift, corresponding to the forcing in the ground model: ${\bf V} \rightarrow {\bf V}[r_A]$, where $r_A$ is Cohen generic real corresponding to A.
\end{theorem}
Proof:  ${\bf V}\simeq{\bf V}^{(B_\alpha )}/U \rightarrow {\bf V}^{((B_\alpha )+A)}/\bar{U}\simeq {\bf V}^{C_\omega}/\bar{U} \simeq {\bf V}[r_a]$ .  $\Box $

{\it Remark:}  In Theorem4 we cannot have both real numbers as a result of measuring non compatible observables, otherwise they would be compatible, and an enlarged the algebra $((B_{\alpha})+(A_s))\simeq C_\omega $ does not appear. We only have algebra $(B_\alpha)$ and consequently standard reals as a result of measuring observables from this algebra.

In that way we have obtained the model of being non compatible observables via forcing construction, in the sense that we are able to show what spectrum could correspond to the "algebra" generated by non compatible observables. Let us note that our forcing involved in the construction is one which changes the meaning of the Boolean Valued universe as in $bBUSuS/U$.

\begin{corollary}
Under $Main$ $Hypo$ the spectrum of the Boolean algebra $(B_\alpha)$ is shifted to the spectrum of enlarged Boolean algebra $((B_\alpha )+(A_s))$ by adjoining some ingeneric reals.
\end{corollary}

Now we are interested in the characterization of the meaning of standardness of real numbers (and standardness of related set--theoretical entities) appearing in measurement of various observables; what are our formal abilities to decide weather we can maintain the notion of standard real over all class of observables? The above  construction allowed us to "see" Nonstandard reals or Ingeneric reals just by referring to well defined notion of Standard Natural or Standard Real. Is it possible to maintain it for all observables and measurements?

It can be shown that this is not the case; the modification of Standard Naturals and reals has to be taken seriously in the context of QM. This means we cannot decide once and for ever about the standardness of sets to "see" (with the bounded Boolean ultrasheaf constructions) the shifts Standard Reals$\rightarrow$ Ingeneric Reals; this is possible only locally.

To close this chapter, let us make a note about some possibilities to consider "Spectral Theorem" associated to the "Boolean algebra", that come from non compatible observables, as in Theorem2. Firstly, there is well developed theory of measure on the internal nonstandard spaces (\opencite{Loeb1982}), so called Loeb measure. Secondly, \inlinecite{Ozawa1994} has extended this measure over ingeneric measure spaces . We can try to characterize "$C^{\star}$-algebra" coming from the above "Boolean algebra" generated by non commuting observables, as the one, which in some sense is isomorphic to the "function algebra" over ingeneric spectrum.  In this way, noncommutativity is eaten by our $Main$ $Hypo$ and the commutative function algebra emerges again.

\section{Model Theoretic Analysis of Small Exotic Smooth Structures on $\mathbb{R}^4$.}  
General references on exotic smoothness topics and Kirby calculus are \inlinecite{GompfStipshitz1999} or \inlinecite{Kirby1989}. There were also attempts to relate exotic smoothness to some physical valid situations (\opencite{Brans}, \opencite{Sladkowski}). 
Let us consider so called small exotic smooth $R^4$.  Small exotic smooth $R^4$'s are all imbedded into Standard $\mathbb{R}^4$. They come from the failure of smooth h-cobordism theorem in dimension 4 (large exotic smooth structures on $\mathbb{R}^4$ come from the failure of  higher dimensional surgery theory applied to dimension 4). There are continuum many different non diffeomorphic smooth structures (small as well large) on topological $\mathbb{R}^4$. Any large smooth $R^4$ contains compact manifold which can not be smoothly imbedded into standard $\mathbb{R}^4$. 
It is known that small exotic smooth $R^4$, as being 4--manifolds, have handle decompositions (infinite) with only 0, 1 and 2--handles included. Typical handlebody of such manifold is given by a compact submanifold of $\mathbb{R}^4$ (so called Akbulut cork) with some Casson handles (CH) attached. CH are infinite towers of kinky handles and any Casson handle is homeomorphic to the standard 2-handle (\opencite{Freedman1982}).
The simplest exotic (small) $R^4$ is represented by the following Kirby diagram (\opencite{BizacaGompf1996})(Figure 1), where it is understood that we are taking only interior of the handlebody ignoring the boundary (except the attaching circle of the CH). Different exotic small $R^4$ differ among themselves just by the complexity of CH and by compact $K$.
\medskip
\begin{figure}[H]
\begin{center}
\leavevmode
\includegraphics[scale=.70]{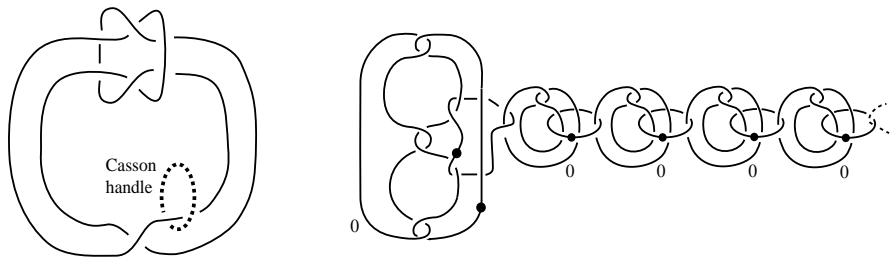}

\caption{Casson handle in the simplest exotic $R^4$.}
\label{figref}
\end{center}
\end{figure}
The motivation to consider small exotic $R^4$ is in that, they might be represented (in principle) by the explicit figures with their Kirby diagrams. 
The idea to use Model Theory to analyze exotic structures on $\mathbb{R}^4$ is in our ability to introduce specific partial order in the set of pairs $(K,{CH_i}^{(n)})$ of compacts $K$ and finite stages of Casson Handless (attached to $K$ to obtain exotic smooth $R^4$).
The partial order in question is to be separative one. Then we can produce boolean algebra of regular open (RO) subsets of the above partial order $P$. So, we have $RO(P)$ (\opencite{Jech}). It is known that this algebra is complete atomless BA (\opencite{Jech}). Taking Ground Model $V$ (as usual this is to be Countable Transitive Model of ZFC to allow forcing), we can create Boolean valued model of ZFC -- $V^{RO(P)}$. Based on this, we are able to retrieve two valued models just by taking quotients by non principal ultrafilters in $RO(P)$. On the level of Ground Model this is just the same as adding some Cohen generic real $r$: $V \rightarrow V[r]$.

Now we can formulate:
\begin{theorem}
Forcing which adds generic Cohen real $r$ to $V$ corresponds exactly to attaching some Casson Handle to K, corresponding in turn to the change of the smooth differential structure on $\mathbb{R}^4$.
\end{theorem}

Next, we are ready to describe how the $Main$ $Hypo$ is going to generate Smooth Exotic Structures on $\mathbb{R}^4$.
\begin{lemma}
Under Main Hypo one can have indistinguishability of some Standard $\mathbb{R}^4$ and some Non Standard ${\bf^\star R^4}: \mathbb{R}^4 \rel {\bf^\star R^4}$.  
\end{lemma}

\begin{lemma}
Dirac $\delta$--function (distribution) on $\mathbb{R}^4$, is smooth ordinary function (with values in ${\bf ^\star R}$) on ${\bf   ^\star R^4}$.  
\end{lemma}

Proof: see \inlinecite{Robinson1979}.

Now, let us observe, that having $\mathbb{R}^4 \rel {\bf   ^\star R^4}$ we do not distinguish formally between smooth functions on ${\bf   ^\star R^4}$ and smooth functions on $\mathbb{R}^4$. This means we should have some smooth function on $\mathbb{R}^4$ which corresponds (not in a direct way) to $\delta$--function on $\mathbb{R}^4$. The change in the smoothness which guarantees the change $\delta$--function $\longrightarrow$ smooth function, is, of course, the same as  the changing standard smoothness into some other one (nondiffeomorphic). This is Non Standard Exotic Smooth Structure.

We have reached the point where Model Theoretical analysis of exotic smooth structures on $\mathbb{R}^4$ shows connections to Set Theoretical forcing and \textit{Main Hypo}. Is it accidental, or does it express any essential correlations between Exotic Smooth $R^4$ and Model Theoretical shape of QM (which we have exhibited in the first part)? We claim that the former is true, but detailed analysis will appear elsewhere.

In what follows we only give an example concerning this situation in the context of the so called $AdS/CFT$ correspondence in $Sustring$ or M--theory.  

\section{Model Theoretic Analysis of the $AdS/CFT$ correspondence -- an example.}
Since \inlinecite{Maldacena1997} made the conjecture about strict mathematical equivalence (duality) of completely different theories: Superstring Theory on $AdS_5 \times S^5$ background and Supersymmetric Conformal SU(N) Yang Mills Theory in 4--dimensions in the limit of large N, much work has been done to check it. There were also some proposals (\opencite{Witten1999}) to make  connection of this duality to "realistic" non supersymmetric, non conformal YM theory in 4-dim (QCD?).

Based on some Model Theoretic analysis of the solutions of 10--dim Supergravity equations we present possible mechanism to generate the sources of masses terms in 4--dim Super YM CFT which might (in principle) break SuSy.

The crucial ingredient of $AdS/CFT$ correspondence is to take N D3--branes as a solutions of classical SuGry equations in 10--dimensions (which is low energy limit of IIB SuString Theory and D3--branes are also present here). Taking N to infinity and considering N D3-branes as coincident, it is possible to recover the near horizon geometry of stack of D3--branes so obtained . This geometry is just $AdS_5\times S^5$. \inlinecite{Witten1998} has shown how the correlation functions in the bulk SuString Theory correspond to the ones in SYM CFT on the 4--dim boundary of $AdS_5$. He has also shown what is the shape of the sources in boundary theory (coming from the bulk) and to what boundary operators they couple. We don't present calculations here, we just state main ideas about how one might generate sources of gravity which would break SuSy in the bulk YM theory.

Let us consider the stack of N D3--branes as a multiplicity of N 4--dim world volumes which is (topologically) a multiplicity of $\mathbb{R}^4$. Let us note that every D3--brane is the solution of 10--dim SuGry. This can be seen as a kind of model of the theory given by 10--dim SuGry equations. We are going to produce another model (Nonstandard) using technique of Ultroproduct (modulo some ultrafilter in $P(\omega )$). We don't discuss here the subtleties concerning orders of the formal languages involved. In such a way we will generate Nonstandard solution to 10--dim SuGry whose world volume is just $(\mathbb{R}^4)^{N \rightarrow \infty}/U$ which is ${\bf  ^\star R^4}$.
Now we are able to state:
\begin{theorem}
Under Main Hypo the procedure described above gives sources of gravity in 4--dim.
\end{theorem}

The proof of this Theorem is based on the analysis in the previous paragraph, where the shift $\delta $--function $\rightarrow$ smooth function on $\mathbb{R}^4$, corresponded to the change of differential structure on $\mathbb{R}^4$, and we know from the work of \inlinecite{Asselmeyer}, how the change of smooth differential structure on compact 4--Manifold generates sources of 4--dim Gravity. 
The possibility of breaking SuSy by these sources comes from analysis of Polchinski and Strassler (\opencite{Polchinski}). The details are to be presented elsewhere.

\section{Summary}
We have presented, very in a sketchy way, the ideology of analyzing QM via Model Theory based on $Main$ $Hypo$. Even more sketchily we have noticed about similar analysis of the exotic (small) smooth structures on $\mathbb{R}^4$. There are striking similarities to both approaches: they require $Main$ $Hypo$ and  a set theoretical forcing appears unexpectedly, yet rather naturally. $Main$ $Hypo$ and forcing are able to code formally the structure of QM in the language of operator algebras over separable Hilbert space. This coding can be used to explain certain aspect of Maldacena conjecture that quantum field theory without gravity is to be dual to the theory describing gravity.

There are also some possibilities to break SuSy or conformal invariance in 4--dim YM Theory in order to obtain connections to realistic (confined) 4--dim QCD.

\end{article}

\end{document}